# Information surfaces and overflow fields: a unitary evolution model based on the principle of information curvature scaling

Zheng Yahui[(a)1,2]

1:Department of Physics, School of Physics and Electronic Engineering,

Sichuan University of Science & Engineering, Zigong, 643000, Sichuan, China

2:Research Center for Ray Detection Discipline and Technology, Sichuan University of Science & Engineering, Zigong, 643000, Sichuan, China

## Abstract

The black hole information paradox exposes a fundamental contradiction between quantum unitarity and semiclassical gravity. Although the entanglement island paradigm is capable of reproducing the Page curve, the physical essence of quantum extremal surfaces and the microscopic mechanism of information transfer have yet to be clarified. Taking quantum information as a fundamental conserved quantity and adopting the principle of information curvature scaling, this paper argues that no singularity exists inside a black hole. Instead, there exist information surfaces encoding quantum information and attached overflow field structures. The model divides black hole evaporation into two stages: the unsaturated stage and the saturated stage. During the unsaturated stage, Hawking radiation dissipates the overflow field, resulting in a steady increase of entanglement entropy. In the saturated stage, the overflow field is completely exhausted. The information surface then undergoes fragmentation and reconstruction, emitting high-energy particles carrying quantum information and driving a gradual decline in entanglement entropy. At the very final stage of black hole evaporation, all residual quantum information and mass are released entirely through a quantum outburst. This model establishes the physical correspondences between the event horizon and the quantum extremal surface, as well as between the information surface and the entanglement island, which can provide some phenomenological constraints for quantum gravity theories.

**Keywords: Information surfaces; Overflow fields; Entanglement islands; Page curves; Planck grains**

### 1. Introduction

The black hole information paradox is a long-standing open problem in quantum gravity and black hole physics, rooted in an underlying fundamental conflict between general relativity and the unitarity and information conservation of quantum mechanics. In 1976, Hawking put forward the black hole evaporation model within the framework of semiclassical gravity, demonstrating that Hawking radiation takes the form of information-free thermal states, implying all quantum information carried by matter falling into the black hole would be permanently erased upon complete black hole evaporation [1]. This result directly contradicts the core tenets of quantum mechanics—the unitary evolution of quantum states and information conservation—and has triggered decades of intensive debate across theoretical physics.

Various theoretical schemes have been proposed by the scientific community to resolve the black hole information paradox. Maldacena et al. proposed the AdS/CFT correspondence [2], which equates gravitational systems in asymptotically Anti-de Sitter (AdS) spacetime with conformal field theories (CFTs) defined on the boundary. The duality inherently enforces unitarity and justifies the equivalence between the RT/HRT area law and the Bekenstein-Hawking area-entropy law; nevertheless, it fails to elucidate the microscopic transfer mechanism of quantum information. By positing a high-energy quantum firewall adjacent to the event horizon, Almheiri et al.'s firewall conjecture [3] severs the entanglement between virtual and real particle pairs

(a) Corresponding author: zhengyahui1979@163.com

associated with Hawking radiation to prevent irreversible information loss, at the cost of invalidating the equivalence principle locally in spacetime.

Maldacena's ER=EPR conjecture (wormhole-entanglement equivalence theory) [4] interprets every pair of entangled particles as being linked by Planck-scale wormholes, replacing dynamical quantum correlations with geometric connectivity to resolve the monogamy conflict underlying the firewall paradox. Nonetheless, this framework leaves the storage and transport of quantum information unaddressed. The quantum extremal surface (QES) [5] and the island paradigm [6] reconcile semiclassical gravity with the Page curve [7] via the definition of generalized entropy, yet they fail to furnish a microscopic dynamical picture for information transfer or account for the physical origin of entanglement islands. Proposals including soft hair, edge-mode and grey-hole scenarios [8-10] postulate that all information is encoded within two-dimensional symmetric structures localized near the event horizon to evade singularity-induced information loss. These approaches are essentially static boundary ansatze rather than full spacetime dynamical descriptions, as they generally lack a dynamical account mapping three-dimensional collapsing matter onto two-dimensional surface degrees of freedom, alongside missing formulations for bulk field evolution, information transport and structural stability.

To tackle the aforementioned predicaments, this paper departs from the conventional research paradigm of patching loopholes for the paradox. Starting from conceptual reconstruction and ontological foundation establishment, we establish a three-tier theoretical framework to develop an original resolution to the issue of black hole information paradox. At the conceptual level, we systematically distinguish six entropy definitions, namely the quantum information entropy, von Neumann entropy, entanglement entropy, coherence entropy, the Hawking entropy and the thermodynamic entropy, and clarify their respective similarities and discrepancies. We argue that the black hole information paradox originates from cross-scale confusion between information carriers and evolutionary laws.

At the fundamental principle level, starting from the unitarity of quantum mechanics, we establish a causal chain ranging from information conservation and microscopic reversibility to quantum unitarity. Information conservation is elevated from an auxiliary constraint of quantum mechanics to a first principle governing the evolution of the microscopic universe, laying the fundamental groundwork for analyzing the unitary evolution of black holes.

At the dynamical level, we introduce the concept of the information surface (IS). Combined with the principle of information curvature scaling (PICS), a two-stage model for black hole evaporation is constructed. Relying on overflow fields and the fragmentation-reconstruction mechanism of information surfaces, we characterize the confinement and release of quantum degrees of freedom throughout black hole evaporation, reinterpret the rising and falling behaviors of the Page curve, and resolve the contradiction between the information-free Hawking radiation and the quantum unitarity.

The remainder of this paper is organized as follows. Section 2 systematically compares and differentiates various entropy concepts, formulates the first principle of information conservation, and proves the intrinsic consistency among the information conservation, microscopic reversibility and quantum unitarity. Section 3 introduces the information surface and principle of information curvature scaling to reinterpret the physical mechanism behind the Page curve, thereby providing a realization route for information conservation. Section 4 establishes a two-stage evolutionary model consisting of unsaturated and saturated black hole evaporation, elucidates the quantum information transfer mechanisms responsible for the rising and falling branches of the Page curve, and endows quantum extremal surface (QES) with concrete physical interpretations.

Section 5 discusses the structure of the information surface. It points out that the packing factor of Planck grains is proportional to the mass of the information surface. The information surface possesses a surface potential barrier that confines abundant quantum foam, and its equivalent pressure acts as a key factor maintaining the stability of the information surface. The section 6 addresses the ultimate fate of black hole evaporation. It demonstrates that the widespread existence of primordial black hole evaporation remnants is inconsistent with astronomical observations and cosmological evolution, and proposes a specific dynamical mechanism for the remnant-free evaporation of black holes.

Section 7 analyzes the evolutionary stages of present-day black holes, concluding that most of them are in the unsaturated stage. Accordingly, the high-energy outbursts accompanied by the fragmentation and reconstruction of the information surface are astronomically unobservable.

Section 8 presents the conclusions and discussions, which systematically summarize the content and innovations of this work.

## 2. The discrimination of entropy concepts

The black hole information paradox originates from the confused usage of entropy concepts: quantum information, von Neumann entropy, entanglement entropy, coherence entropy, Hawking entropy and the thermodynamic entropy. In this section we shall reinterpret the physical connotation, evolutionary law and applicable scope of the above concepts, and analyze the black hole information paradox from its fundamental origin.

### 2.1 The quantum information and quantum information entropy

In this work, quantum information is identified as the intrinsic quantum degrees of freedom of the black hole system, defined as the logarithm of Hilbert-space dimension $d$ [11], namely

$$I = \log_2 d = \frac{S_I}{k_B \ln 2}, \quad (1)$$

where $S_I$ denotes the quantum information entropy of the black hole and the $k_B$ stands for the Boltzmann constant. Although quantum information serves as an equivalent characterization of quantum degrees of freedom, it constitutes a fundamental conserved quantity. Analogous to the conservation of energy and momentum, its conservation is linked to a fundamental physical symmetry: time-reversal symmetry [7]. Based on the premise that the ontology of quantum information originates from the dimensionality of Hilbert space, in present work we propose that the conservation of quantum information is a first-principle law, serving as the cause rather than the consequence of time-reversal symmetry of spacetime. Quantum unitarity and microscopic reversibility both emerge as the direct consequences of this underlying symmetry.

Accordingly, the total quantum information of an isolated quantum system remains invariant. Microscopic reversibility forbids entropy production at the microscopic scale, implying quantum information entropy can only flow spatially from one subsystem to another instead of being spontaneously generated or annihilated out of nowhere. Any dynamical evolution of a quantum system merely rearranges the configuration of its quantum degrees of freedom without altering the total amount of contained quantum information. The quantum information entropy $S_I$, used to quantify quantum information, decomposes into two components: the von Neumann entropy and the coherence entropy, namely,

$$S_I = S_{Vn}(\rho) + S_c(\rho). \quad (2)$$

The von Neumann entropy $S_{Vn}(\rho)$ quantifies the mixing degree of the instantaneous quantum state for a quantum system and is uniquely determined by the system's density matrix. Let $\rho$ be the density matrix of the black hole, and then its von Neumann entropy is defined as

$$S_{Vn}(\rho) = -Tr(\rho \ln \rho). \quad (3)$$

Coherence entropy $S_c(\rho)$ characterizes the coherence magnitude of quantum information inside black holes, which is defined in this work as the difference between quantum information entropy and von Neumann entropy. Defined as the residual component of quantum information entropy herein, this coherence entropy reflects the inherent coherence reserve of black holes. In contrast, the conventional coherence entropy such as relative-entropy-based coherence [12] serves as a state-dependent measure of quantum superposition. Despite their complementary analytical roles, these two types of coherence entropy differ fundamentally in physical interpretation.

The essence of unitary evolution for quantum systems lies in the absence of microscopic entropy production; only quantum entropy flow and rearrangement of quantum configurations take place. Restricted by unitary dynamics, both von Neumann entropy and coherence entropy of an isolated quantum system are conserved, corresponding to reversible microscopic evolution. It is critical to emphasize that von Neumann entropy quantifies the mixing extent of instantaneous quantum configurations rather than the quantum information itself.

### 2.2 The entanglement entropy between subsystems

The entanglement entropy is an observable entropy emerging after partitioning a composite quantum system into two subsystems, which quantifies the quantum correlation strength between the two subsystems. The variations in entanglement entropy do not necessarily indicate the increase or decrease in the total quantum information of the composite system. In other words, the entanglement entropy of a bipartite system is observation-dependent; modifications to subsystem partitioning schemes or the evolution of correlation patterns can both induce changes in the entanglement entropy.

In accordance with Page curve theory, a black hole together with the ambient radiative field can be treated as an isolated pure-state system. Let $|\Psi\rangle_{AB}$ denote the state vector of this bipartite pure system, where subsystem $A$ stands for the black hole and subsystem $B$ corresponds to the radiative field. The reduced density matrix of the black hole is given by

$$\rho_A = Tr_B |\Psi\rangle\langle\Psi|. \tag{4}$$

Since the composite system remains globally in a pure state, the von Neumann entropy of the black hole equals the entanglement entropy between the two subsystems [7], namely

$$S_{VN}(\rho_A) = S_{en}. \tag{5}$$

Page's theory argues that the black hole evolves from an initial pure state to a final pure state throughout the entire evaporation process, with its entanglement entropy starting and ending at zero. Accordingly, no information paradox arises, and black hole dynamics complies with quantum unitarity.

This issue can be further reanalyzed from the perspective of quantum information conservation. Combining Eq. (2) with Eq. (5), we obtain

$$S_I = S_{en} + S_c(\rho). \tag{6}$$

According to the above formula (6), during the early and intermediate stages of the black hole evaporation, Hawking radiation enhances the entanglement between the radiative field and the black hole, giving rise to a growing entanglement entropy of the black hole. The growth of entanglement entropy dissipates the coherent reserve of the internal degrees of freedom of the black hole, the coherence weakens accordingly, and thus the coherence entropy of the black hole decreases monotonically until it vanishes. At this turning point, the entanglement entropy peaks and becomes identical to the black hole's total quantum information entropy. In the late evaporation phase, the available degrees of freedom for entanglement shrink alongside the depletion of the black hole's internal quantum information, which drives a synchronous drop of entanglement entropy. Upon complete evaporation of the black hole, both its entanglement entropy and quantum information entropy converge to zero.

The evolutionary process of black holes is free of the information paradox, consistent with quantum unitarity, and precludes the spontaneous creation or annihilation of quantum information. The early-to-middle evolution corresponds to unsaturated evaporation, whereas the late-stage evolution falls into the saturated-evaporation regime. The detailed characteristics and evolutionary trajectories will be elaborated in subsequent sections.

**2.3 The thermodynamic entropy and Hawking entropy**

The thermodynamic entropy quantifies the number of physically accessible microscopic configurations of a black hole under macroscopic constraints, characterizing the disorder of macroscopic systems and obeying the second law of thermodynamics (entropy increase principle). The underlying origin of thermodynamic entropy growth lies in irreversible processes and local entropy production within the system. From an emergent perspective, rising thermodynamic entropy originates from cross-scale coarse-graining effects. From an observational viewpoint, the entropy increase stems from the information masking of the underlying microscopic quantum configurations, which differs fundamentally from the reduced-density-matrix measurement in the entanglement correlations.

Thermodynamic irreversibility and the growth of thermodynamic entropy are macroscopic emergent effects, which are compatible with unitary evolution and quantum information conservation at the microscopic scale. The rise in macroscopic disorder essentially arises from the transformation of microscopically encoded quantum information from macroscopically observable states into hidden microscopic configurations; no quantum information is lost, and neither microscopic reversibility nor quantum unitarity is violated.

In accordance with general relativity, the area of a black hole's event horizon never decreases, a result known as the area theorem. This allows the definition of an apparent entropy proportional to the horizon area, namely the Hawking entropy [13]. Within the framework of semiclassical gravity, black holes undergo evaporation via Hawking radiation. Black hole evaporation is both a dynamical and a thermodynamic process. The accompanying Hawking radiation elevates the thermodynamic entropy of the radiation field while diminishing the black hole's horizon area, such that the total thermodynamic entropy of the composite system comprising the black hole and surrounding spacetime tends to increase.

In this sense, Hawking entropy can be regarded as the thermodynamic entropy of black holes. That is to say, Hawking entropy is no longer merely an apparent entropy proportional to the horizon area; it is also linked to the macroscopic irreversible evolution of the black hole itself and serves as the thermodynamic entropy characterizing the disorder of the black hole system. Accordingly, the area law for Hawking entropy bears physical reality, which states that the thermodynamic Hawking entropy of a black hole is proportional to the area of its event horizon.

From a microscopic perspective, Hawking radiation carries part of the black hole's quantum information. As Hawking particles are emitted, the entanglement entropy between the radiation field and the black hole first rises and then falls, consistent with the predictions of Page curve theory.

**3. The principle of information curvature scaling**

In the present work, we propose the principle of information curvature scaling. This principle posits that spacetime singularities do not exist inside black holes; instead, a dynamical inner interface emerges where most of the black hole's quantum information is fixed and encoded, defined as the Information Surface (IS). In the black hole interior, the spacetime curvature outside the information surface is proportional to the encoded quantum information and inversely proportional to the cube of the characteristic distance, formulated as

$$\kappa = \xi \frac{I}{r^3}, \tag{7}$$

where $\xi$ denotes an undetermined constant and $r$ stands for the distance from a spacetime point outside the information surface yet inside the black hole to the central core of the black hole. Here $\kappa$ refers to the tidal curvature defined within general relativity. Equation (7) does not contradict general relativity but extends its applicable scope.

The black-hole information surface encodes a fixed quantity of quantum information alongside a corresponding amount of black-hole mass. Their proportionality coefficient $\chi$, termed the information-mass ratio, is a universal constant. Two alternative configurations are available for the arrangement of quantum information on the information surface: close packing and loose packing. The close-packing ansatz directly yields an area law, whereby quantum information entropy scales linearly with the area of the information surface. Nevertheless, this deduction restricts the scope of application of the present model. Accordingly, we discard the close-packing assumption and adopt the loose-packing configuration for quantum information distributed over the information surface. Relevant detailed analysis will be presented in subsequent sections.

The black hole horizon radius is denoted by $r_s$, and the radius of information surface is denoted by $r_a$. For stationary black holes, the relation $r_a<r_s$ generally holds. Black hole evaporation proceeds in two distinct stages: the unsaturated stage and the saturated stage. Prior to saturation, the event horizon and information surface remain spatially separated; upon saturation, these two surfaces coincide and the black hole follows a sophisticated evolutionary trajectory. Saturation is defined as the state, at which maximal entanglement between the interior and exterior of the black hole is realized, once interlayer overflow field is exhausted via Hawking radiation. This two-stage evaporation model shall be elaborated thoroughly in the following section.

**4. The two-stage evaporation model of black holes**

**4.1 The unsaturated stage: information surfaces and overflow fields**

This work restricts its investigation to the evaporation of Schwarzschild black holes. A stationary Schwarzschild black hole forms an isolated pure-state system together with the radiation field residing in the spacetime surrounding its horizon. Black hole evaporation corresponds to the internal redistribution of quantum information within this pure-state composite system, where less quantum information flows from the black hole subsystem into the external radiation field in the unsaturated stage, and massive quantum information transfer takes place between the two subsystems throughout the saturated stage. Evaporation of Schwarzschild black holes is powered by quantum vacuum fluctuations localized near the event horizon, and the tidal spacetime curvature around the horizon is governed by the black hole's mass, that is,

$$\kappa \equiv \frac{1}{r_s^2} \sim \frac{1}{M^2}. \tag{8}$$

The above formula is derived from general relativity. Since quantum information is approximately proportional to black hole mass, this expression is consistent with the information curvature scaling relation Eq. (7).

The spacetime curvature near the horizon triggers asymmetric vacuum fluctuations and generates virtual particle pairs. One virtual constituent falls into the black hole while its real counterpart radiates outward to infinity, which effectively reduces the black hole mass; this physical process corresponds to Hawking radiation. From the perspective of quantum field theory, real and virtual particles are inherently quantum-mechanically entangled. After infalling virtual particles amalgamate with the interior degrees of freedom of the black hole, entanglement is established between outgoing real Hawking quanta and the black hole's internal degrees of freedom. The specific evolution law of entanglement entropy needs to be analyzed based on the structure of information surfaces and overflow fields inside black holes.

The principle of information curvature scaling rules out spacetime singularities inside black holes. Dimensional compactification occurring during gravitational collapse compresses the black hole's three-dimensional degrees of freedom into a closed two-dimensional dynamical surface named the information surface, which constitutes the minimum-energy surface in the black hole interior. A topological phase transition emerges in the course of gravitational collapse, altering the direction of gravitational contraction: contrary to the conventional gravitational theory claiming collapse converges toward the central singularity, infall is directed toward the information surface. Specifically, matter located either inside or outside the information surface radius $r_a$ accretes onto this surface.

Additionally, the black hole possesses a causal boundary, the event horizon. Generated from disparate physical origins, the event horizon and information surface cannot coincide inherently. An event horizon nested within the information surface would violate causal laws, so the horizon is inevitably situated outside the information surface. Analogous to Penrose's cosmic censorship conjecture [14], a causal censorship mechanism prevents the information surface from being exposed beyond the event horizon, which is still termed the cosmic censorship conjecture in this work.

We now proceed to analyze the structural and dynamical properties of the information surface. Intense intrinsic gravity of the black hole induces localized spacetime warping. To sustain structural stability, spacetime dynamics triggers a minimal-scale response characterized by the Planck scale. Upon activation of this Planck-scale dynamical response, local spacetime curvature saturates at its limiting value and spreads uniformly throughout the compressed region; this limiting curvature is defined as the Planck curvature. The uniform spreading signifies that the local spacetime curvature everywhere within the compressed region reaches the extremal magnitude at the Planck scale.

The resultant physical result is the fragmentation of the compressed region into the discrete Planck grains. Such a dynamical response of spacetime corresponds to gravitational quantization. In the present context, the gravitational quantization refers to the quantization of spacetime dynamics rather than spacetime geometry itself. This implies that large-scale macroscopic spacetime geometry remains governed by the continuous geometric laws of classical general relativity, while only the spacetime dynamics associated with microscopic matter inside the compressed region undergoes quantization at the Planck scale.

Each Planck grain has a size equal to the Planck scale $l_P$ and a mass equal to the Planck mass $m_P$, which are expressed as follows

$$l_P = \sqrt{\frac{G\hbar}{c^3}}, \qquad m_P = \sqrt{\frac{\hbar c}{G}} \quad . \tag{9}$$

The Planck curvature is then defined as

$$\kappa_P = \frac{1}{l_P^2} \quad . \tag{10}$$

Planck curvature essentially corresponds to the spacetime tidal curvature on the surface of an individual Planck grain and represents the maximum curvature permitted by spacetime dynamics. Any curvature exceeding this threshold would imply constituent scales falling below the Planck length, where spacetime dynamics can no longer sustain a physical response and known physical laws break down. Quantum-gravitational vacuum fluctuations at the Planck scale spontaneously generate quantum foam [15], which envelops every Planck grain. Acting as an effective shielding medium, quantum foam confines the high-curvature spacetime field within localized domains. Consequently, spacetime regions beyond the reach of quantum foam retain their conventional classical geometry and remain insulated from the intense curvature fields originating from Planck grains.

The spacetime curvature on the surface of a single Planck grain is determined by Eq. (10), from which the proportionality constant in the information curvature scaling relation Eq. (7) can be fixed. Let the information content carried by an individual Planck grain be $I_0$; combining Eq. (7) and Eq. (10) yields

$$\frac{l_P}{l_P^3} = \xi \frac{I_0}{l_P^3} \quad . \tag{11}$$

Hence we obtain

$$\xi = \frac{l_P}{I_0}, \qquad \kappa = \frac{l_P}{I_0} \frac{I}{r^3} \quad . \tag{12}$$

The information surface of a black hole is composed of numerous Planck grains, so the information surface has a finite thickness equal to the Planck scale rather than vanishing thickness. Each Planck grain carries the fixed mass and quantum information, yielding a constant $\chi$ for its information-to-mass ratio. This constant $\chi$ holds exclusively for the interior of black holes; ordinary particles residing in spacetime outside the black hole possess an information-to-mass ratio substantially smaller than that of Planck grains.

This implies that a fraction of the infalling mass (bound with a modest quantity of quantum information) is extruded out of the information surface during black hole collapse. Subject to mass conservation, the expelled mass together with its associated sparse quantum information is confined within the interlayer between the event horizon and the information surface. Accordingly, excess mass persists over the radial range $r_a < r < r_s$ and manifests in the form of a field, named the overflow field. Let the total rest mass of the black hole be $M$ and the mass of its information surface be $M_a$; the equivalent mass of the overflow field is then written as

$$\Delta M = M - M_a \, . \tag{13}$$

The energy-momentum tensor of the mentioned overflow field acts as a low-information-density source within the semiclassical Einstein field equations.

In the early evaporation stage of a Schwarzschild black hole, the ambient radiative spacetime field does not carry any quantum information originating from the black hole, resulting in zero entanglement between the radiation field and the black hole, namely $S_{en}=0$. At this evolutionary stage, the black hole's quantum informational entropy exists predominantly in form of coherent entropy, i.e.,

$$S_I \approx S_c(\rho) \, . \tag{14}$$

The overflow field is strongly decohered with vanishing coherent entropy, whereas the information surface is intrinsically highly coherent and constitutes the primary carrier of the black hole's coherent entropy.

It should be noted that intrinsic entanglement persists between the overflow field and the information surface, and the corresponding entanglement entropy is concealed inside the black hole and excluded from Eq. (14). This originates from our treatment of the black hole as an undivided holistic system: no bipartite partition or reduced density matrix is constructed, such that the intrinsic entanglement entropy remains unobservable and is not incorporated into Eq. (14).

Virtual particles falling into the black hole first couple with excited-state particles of the overflow field and transfer a small amount of quantum information into the surrounding spacetime radiation field, generating entanglement between the radiation field and the overflow field inside the black hole. Owing to the intrinsic entanglement linking the overflow field and the information surface, the entanglement formed between the radiation field and the overflow field is equivalently converted into entanglement between the radiation field and the information surface. This evolutionary process ultimately produces entanglement entropy shared by the radiation field and the information surface.

As Hawking radiation proceeds, the entanglement entropy between the radiation field and the information surface rises progressively, accompanied by a decline in the coherent entropy of the overflow field/information surface. Globally, the coherent entropy of the black hole is gradually transformed into the entanglement entropy between outgoing radiation and the black hole, with the sum of coherent entropy and entanglement entropy approximately conserved.

The above transformation continues until the overflow field confined in the intermediate layer is fully depleted via Hawking radiation, and the event horizon gradually draws closer to the information surface. At this stage, the black hole's coherent entropy is exhausted, and the entanglement entropy between the radiation field and the black hole (information surface) peaks. This point corresponds exactly to the inflection point of the Page curve, marking the transition from the unsaturated to the saturated evaporation stage of black holes.

It should nevertheless be noted that the stability of the information surface is sustained by the pressure originating from the energy-momentum tensor of the overflow field. As the overflow field becomes progressively depleted and the event horizon gradually approaches the information surface, the inward pressure declines steadily, rendering the information surface progressively unstable. We shall address this issue in the subsequent subsection.

**4.2 The saturated Page inflection point: asymptotic dynamical process**

As the overflow field within the intermediate layer is gradually exhausted, the black hole's event horizon continuously moves toward the information surface, which gradually turns unstable. Once the event horizon touches the quantum-foam region surrounding the information surface, vacuum fluctuations can probabilistically shift segments of the information surface into the spacetime exterior to the event horizon. Meanwhile, the cosmic censorship principle forbids the naked exposure of the information surface. Jointly, these two mechanisms trigger partial fragmentation of the information surface.

Local fragmentation of the information surface is accompanied by the emission of high-information, high-energy photons and ejection of a small fraction of baryons. The information-mass ratio of such high-information high-energy photons markedly exceeds that of Planck grains, namely

$$\chi_{\gamma} \gg \chi_{P}. \tag{15}$$

By contrast, the information-mass ratio of baryons is approximately equal to that of Planck grains. In other words, fragmentation of the information surface induces substantial transfer of quantum information from the black hole to its outside spacetime, alongside a reduction in the black hole's entanglement entropy.

Following fragmentation, the information surface rapidly reconstructs into a two-dimensional curved surface under quantum-gravitational stress, accompanied by a moderate inward shrinkage of its radius. The outflow of high-information high-energy photons regenerates the overflow field within the interlayer between the event horizon and the information surface. The newly formed overflow field restores inward pressure from vacuum fluctuations and stabilizes the information surface once again. Nevertheless, the spatial extent of the regenerated overflow field remains minimal such that $(r_s - r_a) \ll r_a$ ; consequently, the event horizon swiftly drifts toward the information surface again and the foregoing dynamical cycle restarts.

Accordingly, the post-saturation evolution of the black hole evaporation corresponds to a dynamical process featuring the recurrent fragmentation and reconstruction of the information surface, accompanied by the emission of high-information high-energy photons and baryon ejection.

**4.3 The post-saturation stage: quasi-stationary evolution**

After evolving to the Page saturation critical point, the black hole will activate the cyclic partial fragmentation-reconstruction mechanism of the information surface. When local regions of the information surface lose stability and fragment, the confined Planck grains detach from the information surface. Upon detachment, Planck grains break free from the confinement of high curvature and disintegrate, producing high-energy photons carrying quantum information as well as a small number of baryons that radiate into the spacetime outside the event horizon. The characteristic timescale of this radiative release is very short, denoted by $t_\gamma$.

Meanwile, the semiclassical Hawking radiation slowly dissipates the energy contained within the overflow field. In the early stage of post-saturation evolution, the black hole possesses a relatively large total mass and low Hawking temperature at the event horizon. The global evaporation timescale $t_f$ of the overflow field greatly exceeds the single-event particle emission timescale, leading to the following temporal relation，

$$t_\gamma \ll t_f \,. \tag{16}$$

This temporal relation ensures that, after each round of local fragmentation and particle emission, the system possesses sufficient time to accomplish the local reconstruction of the information surface. Upon the completion of reconstruction, the information surface contracts inward; residual energy from high-energy emission fills the interlayer space to reconstruct the overflow field. The overflow field is then slowly depleted via Hawking radiation, initiating the next evolutionary cycle.

Further inferences can be drawn from the temporal inequality in Eq. (16): the characteristic timescale governing the macroscopic relaxation evolution of the black hole during the saturation stage is dominated by the slowly dissipative evaporation of the overflow field. Fragmentation and reconstruction of the information surface remain confined to localized curved regions, and full collapse of the entire surface is prohibited by the cosmic censorship conjecture. Owing to the timescale separation featuring transient local fluctuations and slow global variation, the whole evolution of the black hole in its saturation stage qualifies as quasi-steady-state evolution.

As the black hole enters post-saturation quasi-steady evolution, the Page curve has passed its inflection point and begins to decline. At this stage, the quantum information entropy of the black hole approximately equals the entanglement entropy between the radiation field and the black hole's information surface, namely

$$S_I \approx S_{en} \,. \tag{17}$$

As the information surface shrinks, the entanglement entropy between the exterior radiation field and the black hole inevitably decreases synchronously.

This process proceeds continuously until the complete evaporation of the black hole, with all quantum information fully released synchronously. No quantum information is lost throughout the whole evolution, thereby preserving quantum unitarity.

**5. The information surface structure**

The structure of the information surface forms immediately upon the birth of a black hole. During unsaturated stage, the arrangement pattern of Planck grains on the information surface undergoes no fundamental changes. Once entering the saturated stage, the event horizon and the information surface nearly coincide when the black hole is in a quasi-steady state. Regardless of the specific dynamical mechanisms, the two surfaces should shrink inward synchronously; otherwise, the quasi-steady evolutionary features cannot be maintained.

Accordingly, there should be

$$r_s \approx r_a \,. \tag{18}$$

According to general relativity, the radius of a black hole's event horizon satisfies

$$r_s = \frac{2GM_a}{c^2} \ , \tag{19}$$

where $M_a$ denotes the total mass of the information surface. The total number N of Planck grains residing on the information surface is defined as

$$N \equiv \frac{M_a}{m'_P} \ , \tag{20}$$

where $m'_P$ stands for the effective Planck mass. We define a dimensionless loose-packing factor α, to characterize the degree of loose packing of Planck grains on the information surface, yielding the relation

$$A_a = 4\pi r_a^{\ 2} = N\alpha l_P^{\ 2} = \frac{M_a}{m'_P}\alpha l_P^{\ 2} \ . \tag{21}$$

Thus, we have

$$r_a = l_P\sqrt{\frac{\alpha M_a}{4\pi m'_P}} \ . \tag{22}$$

From Eq.(18), one obtains

$$\frac{l_P}{\sqrt{4\pi m'_P}} \approx \frac{2G}{c^2}\sqrt{\frac{M_a}{\alpha}} \ . \tag{23}$$

If the Planck particles adopt a close-packed structure on the information surface, namely $\alpha$=1, or form a loosely arranged structure where the packing factor $\alpha$ is independent of the total mass $M_a$ of the information surface, only black holes with specific masses can satisfy Equation (18). This implies that regardless of the initial mass of a black hole, its mass must decrease to a certain value to meet the above equation, which limits the applicable scope of the model proposed in this paper. A self-consistent evolutionary model for black holes should yield similar evolutionary paths for black holes of all mass spectra. To make our model applicable to black holes across all mass spectra, the optimal approach is to set the right-hand side of Equation (23) to be independent of the black hole mass. In other words, the packing factor $\alpha$ has a simple linear proportional relationship with the information surface mass $M_a$, i.e.,

$$\alpha \approx \frac{16M_a\pi m'_P G^2}{l_P^{\ 2}c^4} = \frac{16M_a\pi m'_P G}{\hbar c} \ . \tag{24}$$

The effective Planck mass $m'_P$ therein requires separate calculation.

From Eqs. (8) and (12), the spacetime tidal curvatures near the event horizon and the information surface can be written respectively as

$$\kappa_s = \frac{1}{r_s^{\ 2}}, \qquad \kappa_a = \frac{l_P}{I_0}\frac{I}{r_a^{\ 3}} \ . \tag{25}$$

As the event horizon gradually approaches the information surface, Equation (18) is satisfied and the spacetime curvatures adjacent to the two surfaces become approximately identical, such that $\kappa_s \approx \kappa_a$. With this approximation relation, the total number $N$ of Planck grains on the information surface can be derived as follows

$$N = \frac{I}{I_0} \approx \frac{r_s}{l_P} = \frac{2GM_a}{c^2}\sqrt{\frac{c^3}{\hbar G}} = \frac{2M_a}{m_P} \ . \tag{26}$$

Comparing the above expression with Equation (20) yields that

$$m'_P \approx \frac{1}{2}m_P \, , \tag{27}$$

This equation indicates that the effective Planck mass of the information surface is lower than its bare mass, which demonstrates that Planck grains exist in a bound state within the information surface. Inside the information surface, namely the region where $r<r_a$ , the spacetime is flat and the effective potential can be set to zero. Qualitatively, the bound state of the information surface corresponds to a negative total effective energy, while the effective potential is positive in the overflow field of the interlayer. As hypothesized previously, the information surface is indeed the interface with the minimum energy and thus the highest stability inside the black hole.

Furthermore, Equation (27) reveals that the effective Planck mass is independent of the total mass of the information surface. This suggests that the binding effect of the information surface primarily arises from the surface state effect of the quantum foam region surrounding it. An equivalent surface potential barrier exists here, preventing Planck grains and quantum foam inside the information surface from detaching and entering the overflow field region.

From Equation (27), we obtain

$$\alpha \approx \frac{8\pi M_a}{m_P} . \tag{28}$$

This equation shows that for a black hole with larger mass, the Planck grains on its information surface are distributed more sparsely. It implies that a large amount of quantum foam is confined within the information surface. Such quantum foam is excited by the strong curvature on the surface of Planck grains, and the total quantity of quantum foam increases approximately linearly with the total mass of the information surface. The repulsive effect generated by quantum foam drives the outward expansion of the information surface, which is described by the equivalent pressure $P_{out}$ . The curvature jump on both sides of the information surface produces an inward spacetime geometric pressure, denoted as $P_G$ . Meanwhile, the energy-momentum tensor of the overflow field also gives rise to an inward pressure $P_\varepsilon$. When the information surface stays in a steady state, these three forces are in balance, namely

$$P_G + P_\varepsilon = P_{out} . \tag{29}$$

After the overflow field in the interlayer is fully depleted, the aforementioned balance is broken. The equivalent repulsive force of quantum foam causes local fragmentation of the information surface. Some Planck grains disintegrate, emitting high-energy photons and a small number of baryons into the spacetime outside the black hole, and regenerating the overflow field within the interlayer. Once the overflow field is restored, the balance is re-established. The slow evaporation of the overflow field driven by Hawking radiation is initiated, and the black hole enters a new cycle of quasi-steady state evolution.

It should be noted that the above theoretical model is a qualitative model derived from reasoning and applicability constraints, which does not involve the dynamical structure of spacetime. Nevertheless, it imposes constraints on potential spacetime dynamics and prospective quantum gravity theories.

Throughout the full course of black hole evaporation, the entanglement entropy rises initially and declines afterward, perfectly reproducing the trend of the Page curve. Beyond consistency with the Page curve, the present qualitative theoretical model establishes a natural logical correspondence with the entanglement island paradigm. The rising segment of the entanglement entropy corresponds to the unsaturated stage of black hole evolution, during which the quantum extremal surface coincides with the event horizon; the falling entropy segment matches the quasi-stationary evaporative stage after black hole saturation, where the quantum extremal surface lies adjacent to the information surface, and nearly overlaps its surface potential barrier. In the physical essence, the information surface serves as the entanglement island or the fundamental physical carrier of the island.

One open problem remains unaddressed within the current framework: what is the ultimate fate of an evaporating black hole? This issue will be elaborated in the subsequent section.

**6. The ultimate fate of evaporation: black hole remnants?**

Based on the quasi-steady evolution theory for saturated black holes, the quantum information and rest mass stored on the information surface are released synchronously, with the release process mediated and coupled by the overflow field confined in the interlayer between the event horizon and information surface. Separated from the information surface by quantum foam, the overflow field has no direct physical contact with the information surface; instead, the two subsystems are linked via intrinsic quantum entanglement.

We demonstrate the inevitability of such intrinsic entanglement from two perspectives. In accordance with the holographic principle, the information surface acts as the physical boundary of the interior black hole spacetime and encodes most of its quantum information, which necessitates inherent quantum entanglement between bulk interior fields and the boundary surface. In addition, the overflow field originates from the extrusion of material off the information surface. Quantum information residing on the information surface maintains strong coherence; even though the overflow field undergoes decoherence, local correlations persist between it and information surface, in the form of weak intrinsic local entanglement.

The existence of this intrinsic entanglement mechanism ensures that the overflow field can continuously perform the mediating and coupling function. Hence, throughout most of the timescale of quasi-steady evolution, the quantum information and mass of the black hole are released via steady-state Hawking radiation. However, in the very late stage of evaporation when the black hole mass becomes extremely small, according to the Hawking temperature formula

$$T_H = \frac{\hbar c^3}{8\pi k_B GM}, \tag{30}$$

its horizon temperature rises sharply and the power of Hawking radiation increases drastically, rendering Equation (16) invalid. In other words, the evaporation timescale of the overflow field becomes shorter than the emission timescale of all types of photons from the information surface, namely

$$t_f < t_\gamma . \tag{31}$$

This implies the fragmentation-reconstruction mechanism of the information surface ceases to function. Before the information surface completes reconstruction, the black hole event horizon has already fallen back into the quantum foam adjacent to the information surface, making the surface unstable. At this point, the cosmic censorship conjecture forces Hawking radiation to shut down.

After the failure of the Hawking radiation mechanism, the black hole evaporation follows two evolutionary paths. First, it fully seals the information surface and evolves into an evaporative remnant to prevent the loss of quantum information. Second, it triggers complete fragmentation of the information surface. Through a one-time intense outburst, all quantum information and mass stored on the surface are released into the external spacetime, also avoiding information loss.

We argue that the first evolutionary path is unfeasible. The hypothesis of black hole remnants contradicts astronomical observations and the overall picture of cosmic evolution. If the remnant scenario were valid, numerous primordial black holes formed in the early Universe after the Big Bang would have left a vast number of remnants by the present day. These remnants cannot be consumed further during stellar evolution. Their gravitational effects would trigger widespread astrophysical instabilities and produce distinguishable signatures detectable by astronomical observations. Nevertheless, no such evidence has been found in current observations. Furthermore, from the perspective of long-term cosmic evolution, these remnants, unable to participate in material cycles, would act as a form of "entropic waste". Their existence would break the self-consistency of the cosmic evolution and reduce the friendliness of cosmological models.

The second evolutionary path is more theoretically and logically natural. When the power of Hawking radiation surges sharply and condition (31) is satisfied, the black hole event horizon contracts rapidly and the overflow field vanishes abruptly, severing coupling and entanglement with the information surface and disabling Hawking radiation. Driven by quantum gravitational effects, the black hole abandons the stable emission mechanism relying on Hawking radiation and the overflow field, leading to the global fragmentation of the information surface. All remaining quantum information and mass are released through a one-time outburst of high-energy photons and baryon ejection, manifesting as a outburst event. Unlike outbursts triggered by intense radiation in classical black holes, this phenomenon arises from quantum gravity. In this process, no quantum information is lost, and quantum unitarity is preserved.

**7. The evolutionary states of black holes in the present universe**

All the analyses above are based on the assumption of isolated pure states, namely that a black hole and its surrounding radiation field form an isolated pure-state system. Under this premise, we first estimate the Page time, i.e., the time required for a Schwarzschild black hole to evolve from its initial state to the Page turning point. Afterwards, we extend the discussion to open systems. For an isolated system, the overflow field of a Schwarzschild black hole evaporates slowly via Hawking radiation, hence we have

$$\frac{d}{dt}\Delta M = -\frac{\Gamma_{out}}{c^2}. \tag{32}$$

Here, $\Gamma_{out}$ denotes the power of Hawking radiation, which is expressed as

$$\Gamma_{out} = \sigma A_s T_H^{\ 4}. \tag{33}$$

The Stefan-Boltzmann constant and the event horizon area are given by

$$\sigma = \frac{\pi^2 k_B^{\ 4}}{60\hbar^3 c^2}, \qquad A_s = 4\pi r_s^{\ 2}. \tag{34}$$

Substitute the total black hole mass $M$ for the mass in Equation (19), and combine it with Equation (29) into Equation (33). We then obtain

$$\Gamma_{out} = \frac{\hbar c^6}{15360\pi G^2 M^2}. \tag{35}$$

During the unsaturated evaporation stage, the mass of the information surface remains constant. According to Equation (13), we have

$$\frac{dM}{dt} = -\frac{a}{M^2}, \tag{36}$$

where there is

$$a = \frac{\hbar c^4}{15360\pi G^2}. \tag{37}$$

Let the initial mass of the black hole be $M_0$. According to Equation (36), the total time required for the black hole to evaporate completely is

$$\tau_{evap} = \frac{1}{3a} M_0^{\ 3} = \frac{5120\pi G^2}{\hbar c^4} M_0^{\ 3}. \tag{38}$$

Assume that at the early stage of black hole evolution, the equivalent mass of the overflow field is half of the black hole mass, i.e.,

$$(\Delta M)_0 \approx \frac{1}{2} M_0. \tag{39}$$

Then the Page time, namely the time taken for a black hole to evaporate half of its mass, is

$$\tau_{Page} = \frac{1}{3a}\left( M_0^{\ 3} - \frac{1}{8} M_0^{\ 3} \right) = \frac{7}{8}\tau_{evap}. \tag{40}$$

It should be clarified that the above definition differs from the conventional “Page time” defined based on the inversion of entanglement entropy. The latter is a concept from the perspective of quantum information, while the present work adopts a more straightforward definition based on the evolutionary node of mass. Equation (40) indicates that the Page time accounts for 7/8 of the total lifetime of a black hole. For a black hole with one solar mass, its total evaporation time is

$$\tau_{evap} \approx 2\times 10^{67}\, yr. \tag{41}$$

The Page time and the total evaporation time are of the same order of magnitude, both far exceeding the current age of the Universe ($10^{10} yr$).

Furthermore, when the effects of accretion disks and cosmic background radiation are taken into account, Equation (32) is rewritten as

$$\frac{d}{dt}(\Delta Mc^2) = \Gamma_{in} - \Gamma_{out} . \tag{42}$$

Here, the $\Gamma_{in}$ denotes the input power from the accretion disk and cosmic background radiation surrounding the black hole. Except for the primordial black holes, all black holes formed via gravitational collapse after the Big Bang satisfy the following conditions due to their open environment

$$\Gamma_{in} > \Gamma_{out} . \tag{43}$$

This implies that the overflow fields of most black holes in the present Universe are far from being depleted. Assuming the equivalent mass of these overflow fields accounts for half of the total black hole mass, we obtain the following results from Equation (22) and Equation (28)

$$\frac{r_s}{r_a} \approx \frac{M}{M_a} \approx 2 . \tag{44}$$

In other words, the event horizon radius of most black holes is approximately twice the radius of their information surface. Given such a vast interlayer space, the overflow field and the information surface exhibit rich dynamical behaviors.

This paper introduces the concept of the information surface and constructs a information surface-overflow field structure and staged evolution model for black holes, aiming to provide a qualitative dynamical interpretation for the Page curve and the entanglement island paradigm. Since a complete set of dynamical equations has not been formulated, quantitative predictions for observational tests cannot yet be obtained. A key prediction of this model is that when a black hole evolves into the saturated stage, the information surface will undergo fragmentation and reconstruction, break through the event horizon, and emit high-energy photons carrying abundant quantum information as well as a small number of baryons outwards. In principle, such signals are observable. Nevertheless, most black holes in the present Universe have not yet reached the saturated stage, so this process theoretically does not occur for them.

Naturally, the above conclusions are drawn based on two fundamental premises. First, the black hole horizon acts as a classical one-way membrane, through which information and matter can only fall inward. The overflow field escapes outward solely via classical Hawking radiation and remains unaffected by the accretion field. Second, the ratio of the equivalent mass of the overflow field to the total mass of the black hole is independent of the black hole's mass and type.

In fact, this ratio may be correlated with the formation pathway of the black hole. Black holes formed by the direct collapse of massive progenitor stars (including primordial seed black holes) skip intermediate evolutionary stages such as white dwarfs and neutron stars. Their collapse proceeds more abruptly, resulting in a relatively larger proportion of the overflow field. In contrast, intermediate-mass black holes that form through successive evolutionary stages tend to have a smaller overflow field fraction. This leads to differences in the Page transition time for black holes originating from distinct formation pathways.

On the other hand, other dynamical processes may exist within the black hole's overflow field. In the unsaturated stage, such processes can break through the causal shielding of the event horizon and emit high-energy radiation that carries no quantum information, which will significantly shorten the Page time. Furthermore, this kind of endogenous high-energy outburst can provide a more self-consistent dynamical explanation for high-energy events observed in black holes of various mass ranges. The detailed internal dynamics are beyond the scope of this paper and will be further investigated in follow-up studies.

## 8. Conclusions and discussions

This paper clarifies conceptual ambiguities existing in the researches on the black hole information paradox and reconstructs its underlying logic. The key to resolving the paradox lies in defining quantum information as the logarithm of the dimensionality of a system's Hilbert space, namely regarding quantum degrees of freedom as the genuine carrier of quantum information conservation. The entanglement entropy of a black hole characterizes quantum correlations between subsystems. The unitary evolution it describes corresponds to the conservation of quantum states, rather than the quantum degrees of freedom. By taking quantum information as a fundamental conserved quantity and treating quantum information conservation as a first principle, this work presents a new perspective for the thorough resolution of the black hole information paradox.

From this new perspective, acting as a subsystem of a composite pure-state system, the black hole's quantum information entropy consists of two components: the entanglement entropy between radiation and the black hole, and the coherence entropy inside the black hole. As a fundamental conserved quantity, the quantum information entropy cannot be created out of nowhere or destroyed; it can only transfer from one region to another. This conservation essentially stems from quantum unitarity and time-reversal invariance. At the microscopic scale, all physical processes are time-reversal symmetric. There is no entropy production, and only the flow of quantum information entropy occurs.

Based on the principle of information curvature scaling, we propose a new spacetime curvature formula (7). This formula does not negate general relativity, but rather extends it. It directly correlates the quantum information on the black hole's information surface with the spacetime curvature near the surface, which is consistent with modern physical understanding of quantum information [16]. This suggests that the spacetime manifold, as a continuum, may emerge from more fundamental physical realities [17], and such underlying realities are likely deeply linked to quantum information.

We do not present a complete theory of quantum gravity. Instead, we conduct self-consistency analysis by introducing the gravitational quantization mechanism and the Planck scale based on the principle of information curvature scaling. There exists no singularity inside black holes [18]. Instead, an information surface is formed here that concentrates enormous quantum information and part of the black hole mass, which corresponds to the surface of minimum internal energy. The local spacetime curvature of this information surface equals the Planck curvature, and its thickness is on the Planck scale. Composed of numerous Planck grains, the information surface is a physically real structure arising from gravitational quantization. Here, gravitational quantization refers not to the quantization of spacetime geometry, but to the quantization of the spacetime dynamical responses. This viewpoint may facilitate the development of a fully self-consistent quantum gravity theory.

The Planck scale is the minimum scale for the dynamical response of spacetime. Below this scale, spacetime manifests purely as a mathematical continuum, yielding no dynamical responses or physical effects, and thus no physical laws hold. The interior of a black hole's information surface, the spacetime region with a radius smaller than that of the information surface, is precisely such a domain free of any dynamical response. This region contains nothing whatsoever: no fields, no particles, no quantum excitations of fields, and no physical reality. The dynamical quantization of spacetime near the information surface gives rise to abundant quantum foam [15]. The equivalent pressure generated by its strong fluctuations acts as the key factor that drives the expansion of the information surface and sustains the quasi-stationary evolution in the saturated stage.

The event horizon of a black hole lies on the outside, while the information surface resides on the inside. An overflow field exists in the interlayer between the two curved surfaces, which contains a small number of quantum degrees of freedom. During the unsaturated stage of black hole evaporation, the virtual particles first couple with the excited particles in the overflow field and subsequently become entangled with the information surface. As Hawking radiation proceeds, the entanglement between the radiation field and the black hole's information surface gradually intensifies, and the entanglement entropy keeps increasing until the overflow field is fully depleted and the entanglement entropy reaches its maximum value. This corresponds to the rising segment of the Page curve.

After the overflow field is depleted, the black hole enters the saturated stage. Black hole saturation means there is no free quantum information within the interlayer, and all quantum information is stored entirely on the information surface. This indicates that the event horizon has shrunk to the vicinity of the information surface. The black hole then undergoes quasi-stationary dynamical evolution characterized by the fragmentation and reconstruction of the information surface. The fragmentation and reconstruction processes of the information surface are accompanied by the emission of high-information and photons of various species as well as a small number of baryons on short time scales. Meanwhile, the overflow field within the interlayer regenerates, which separates the event horizon from the information surface once again. The quasi-stationary evolution of the information surface leads to massive leakage of the quantum information, and a continuous drop in entanglement entropy. This constitutes the descending segment of the Page curve [7].

Upon entering the quasi-steady evolutionary stage after saturation, the high-energy emission from a black hole's information surface becomes an observable phenomenon. Nevertheless, black holes in the present Universe have not yet reached the quasi-steady evaporation stage, so this observable signature does not exist theoretically.

We hold that black holes leave no remnants in their final evolutionary phase. At this stage, the evaporation timescale of the overflow field is shorter than the emission timescale of high-energy photons from the information surface. The rapid inward contraction of the information surface is no longer hindered, leading to the complete dissipation of the overflow field. The conventional Hawking radiation mechanism can no longer sustain the stability of the information surface. Driven by quantum gravitational instabilities, the information surface undergoes full fragmentation. All quantum information and mass contained within the information surface are completely dissipated in a single outburst. Throughout this process, quantum information is not lost but merely transferred.

The structure of information surface and overflow field, and staged evaporation theory of black holes proposed in this work, are inherently correlated with the entanglement island theory, which has been developed in modern physics to resolve the black hole information paradox and interpret the Page curve. They offer a new alternative for the microscopic dynamical interpretation of the entanglement island theory. The logical framework established in this paper presents a novel solution to the black hole information paradox.

As a phenomenological theory, the staged evaporation theory of black holes does not contradict general relativity; instead, it extends general relativity to the strong gravitational field regime of black holes. Meanwhile, it imposes phenomenological constraints on potential new spacetime theories.

**Acknowledgements**

The author thanks colleagues in the Department of Physics at Sichuan University of Science & Engineering for valuable discussions on black hole thermodynamics and elementary particle physics. This work is supported by the National Natural Science Foundation of China under Grant No. 11405092.